\documentclass[twocolumn 10pt, dvipsnames]{autart}    
\usepackage{booktabs,here}
\usepackage{graphicx,amsmath,amssymb,overpic,here,multirow}          
                               
\usepackage{booktabs}

\usepackage{tikz}
\usetikzlibrary{shapes,arrows}
\usetikzlibrary{intersections}
\usepackage{xcolor,bm}
\usepackage{mathtools}

\begin{document}

\begin{frontmatter}

\title{Learning-Based sensitivity analysis and feedback design for drug delivery of mixed therapy of cancer in the presence of high model uncertainties\thanksref{footnoteinfo}} 
\thanks[footnoteinfo]{This paper was not presented at any IFAC meeting. Corresponding
author M.~Alamir. Tel. +33476826326. Fax +33476826388. This work was supported by MIAI @ Grenoble Alpes under
Grant ANR-19-P3IA-0003.}
\author[GIPSA]{Mazen Alamir}\ead{mazen.alamir@gipsa-lab.inpg.fr}    
\address[GIPSA]{\sc Univ. Grenoble Alpes, CNRS, Grenoble INP, GIPSA-lab, 38000 Grenoble.}
\begin{abstract}
In this paper, a methodology is proposed that enables to analyze the sensitivity of the outcome of a therapy to unavoidable high dispersion of the patient specific parameters on one hand and to the choice of the parameters that define the drug delivery feedback strategy on the other hand. More precisely, a method is given that enables to extract and rank the most influent parameters that determine the probability of success/failure of a given feedback therapy for a given set of initial conditions over a cloud of realizations of uncertainties. Moreover predictors of the expectations of the amounts of drugs being used can also be derived. This enables to design an efficient stochastic optimization framework that guarantees safe contraction of the tumor while minimizing a weighted sum of the quantities of the different drugs being used. The framework is illustrated and validated using the example of a mixed therapy of cancer involving three combined drugs namely: a chemotherapy drug, an immunology vaccine and an immunotherapy drug. Finally, in this specific case, it is shown that dash-boards can be built in the 2D-space of the most influent state components that summarize the outcomes' probabilities and the associated drug usage as iso-values curves in the reduced state space. 
\end{abstract}

\end{frontmatter}
\section{Introduction}\label{sec-introduction}
Rationalizing drug delivery in biomedical applications is obviously a feedback control problem showing the standard feedback control keywords, namely: {\bf control objective} (contraction of disease indicators), {\bf constraints satisfaction} (keeping healthy state), {\bf control saturation} (maximum drug delivery) and {\bf robustness} to uncertainties (on the knowledge of the underlying dynamics) and last but not least, the need for {\bf measurement-based feedback} (the level of drug delivery should depend on the current state of the patient as indicated by the on-line measurements) \cite{Heller2005,MurrayI2002,MurrayII2002}.

While this obvious fact impulsed a tremendous amount of work and concrete devices manufacturing in some use-cases such as the Diabetes monitoring and glucose regulation (see \cite{Moscoso2020,Tirado2018,Polonski2015,Doyle2014,Dalla-Man2014} and the references therein), it is a fact that in the case of cancer treatment we are still far from such automatic processing of drug delivery. 

As a matter of fact, the reasons for such a difference is not so obvious. Indeed, both dynamics can be modeled using population models; for both diseases, the real true dynamics is far more complicated for such simple models to be fully satisfactory. Both dynamics are highly dependent on the patient and drastically vary over time even for the same patient. Nevertheless, despite of these serious common issues, such a mathematical model has been accepted in 2008 by the American Food and Drug Administration (FDA) as a substitute for preclinical trials of certain insulin treatments, including closed-loop algorithms \cite{Dalla-Man2014}. In such a framework, each virtual patient is represented by a randomly sampled (through precise statistical rules) vector of model's parameters. This FDA decision was a paradigm changing step that impulsed a huge amount of works leading to hundreds of publications in peer reviewed journals as well as some real-life automatic insuline delivery devices. 

The implicit rationale behind the FDA decision is that an algorithm that ignores the precise values of the virtual patient's parameters and still succeed in achieving the treatment goals on a statistical basis can reasonably be considered as a valid candidate algorithm for real subjects. This inference is supported by the implicit assumption according to which, for each real subject there exists a set of parameters of the population model that captures the specific patient dynamics to a sufficiently convenient level for the certification to hold. The fact that these parameters cannot be precisely identified is no more a serious issue provided that the following two conditions hold: 1) any real subject dynamics belongs to the family of dynamics generated using the population model by spanning the space of possible values of the model's parameters and 2) the proposed algorithm performs statistically well on the family of virtual patients. 

  It is possible that a similar decision will be made sooner or later for models attempting to describe the population dynamics involved in cancer therapy and this justifies the studies based on such uncertain population models such as the one described in this paper. 

The study and the analysis of population models for cancer dynamics under different kinds of therapies is quite an old research track in the applied mathematics community. In particular, a very rich literature concerns the use of open-loop optimal control problems' solutions in order to infer some insight on best therapy patterns. The precise analysis of this very large literature is obviously beyond the scope of this contribution (see \cite{SWAN1990237,MATVEEV2002311,mazensophie2007,Schattler2015} and the references therein). These works consider the parameters of the population model to be perfectly known. Therefore the impact of model's discrepancy and parameters uncertainties on the outcome of the drug's delivery schedule and the associated patterns is not explicitly analyzed nor the robustness of the nominal design is evaluated or taken into account in the definition of optimality. Therefore, this long tradition of open-loop optimal control design ignores the gain that one might obtain when using feedback that monitors the state of the patients and re-schedule the drug delivery accordingly. 

During the last decade, many works appeared that addressed the cancer therapy as a feedback control problem  (see \cite{Chareyron:09,ZUBAIR2021102804,Czako2017,HIRATA2014278,SHARP2020110277} and the references therein). By so doing, the resulting therapy possesses an intrinsic degree of robustness against parameter mismatches as it is the case in almost any feedback framework. However, the robustness is not precisely quantified nor the statistics of the discrepancy is explicitly taken into account in the design step. Moreover, many of the feedback laws invoked in these works assume a continuously applied state-dependent drug injection which does not really fits the protocols as it is explained later on (see Figure \ref{fig_protocol}). 

A first attempt to explicitly take into account the parametric uncertainty has been proposed by  \cite{alamir_robust_cancer:2014} through the solution of the Hamilton-Jacobi-Isaacs (HJI) equations that enables to design a robust feedback that guarantees the fulfillment of some criteria over a whole domain of admissible parameters. This appealing approach comes however with two drawbacks, namely: the non scalability of the HJI-based formulation on one hand and the over-pessimistic nature of the robust design paradigm on the other hand since robust design addresses the worst disturbance scenario regardless of its probability of occurrence. Robust approach has been recently investigated \cite{MoussaACC2020} using the Lasserre hierarchy \cite{Lasserre2010} which unfortunately shows the same limitations in terms of scalability.

In \cite{ALAMIR201559}, a different, more realistic and more scalable paradigm is proposed. More precisely, a probabilistic certification approach is used \cite{alamo2009randomized,alamo2015} in order to choose the parameters of an explicit analytical state feedback so that a safe contraction\footnote{By safe, it is meant that the contraction occurs while the level of circulating lymphocytes remains greater than some safety threshold during the whole therapy.} of the tumor size can be certified to take place at the end of the therapy with a high probability. The accepted level of failures and the confidence with which the certification can be stated determine together the number of realizations of the uncertainty that have to be drawn (according to the probability distribution governing the dispersion of the parameters) and used in the design procedure (see \cite{ALAMIR201559} for a detailed description of the approach). 

The current contribution follows up on the previous track that takes explicitly into account the statistics of the parameters discrepancy while using a problem-dependent explicitly defined feedback law (contrary to Model Predictive Control implicit induced definition). Moreover, the time structured therapy period which is decomposed into successive treatment and rest periods is explicitly respected leading to a more real-life compatible framework (Figure \ref{fig_protocol}). 

Compared to \cite{ALAMIR201559}, the following new features are proposed in the current paper:

(1) \textbf{Global sensitivity estimation}. A method is proposed to study the sensitivity of the outcome of the therapy to both model's parameters and controller's degrees of freedom. Moreover, the method is not based on the linearization around some operating point as this is irrelevant in the cancer treatment. Rather, the method is based on the use of Machine Learning (ML) tools. This is an interesting contribution for its own as it enables to precisely know what is the subset of {\bf model's parameters} that really matter and might be important to identify for a better precision of the prediction. Indeed, commonly used models might involve a high number of parameters (up to 24 for the supporting combined therapy model used in the current contribution) making identifiability from scarce measurement data a priori cumbersome if not literally impossible. But if it can be proved that only few of them (2-3) are really detrimental to the issue of the therapy, it might become reasonable to identify them on line using dynamic optimization for instance \cite{KUHL201171,alamirObs2021} while considering the remaining parameters to be equal to their nominal median values\footnote{Although this is not what is done in the current contribution}. Similarly, identifying the {\bf control parameters} that really matter facilitates the search for the optimal feedback design by viewing those parameters as the only degrees of freedom (decision variables) in the optimization while freezing the remaining parameters to some reasonably chosen predefined values.

(2) \textbf{Derivation of outcome's dashboards}. While the methodology proposed in \cite{ALAMIR201559} applies to each specific initial state, the ML-based methodology proposed in the present paper enables to derive a success/failure probability maps as well as a drugs' usage expectation maps as functions of a so called \textit{feature vector}\footnote{In the wording of Machine Learning.} which includes: the initial state, the model's parameters vector and the controller's parameters vector in one shot. This highly accelerates the post-processing needed for the sensitivity analysis invoked in the previous item but also, it makes easy to span a grid in the state space (the subspace of relevant components as invoked in the previous item) in order to build a whole picture of the outcome (for a given probability) and the associated cost in terms of the expected drug usage. Note that without the global sensitivity analysis that determines the relevant subspace of state components, such dashboard building step would be impossible to build and its low-dimensional representation would be questionable. 

(3) \textbf{Derivation of a state dependent control parameterization}. As a by-side product of the dashboard construction step invoked in the previous item, a state-dependent choice of the parameters of the explicit feedback law can be obtained. This reduces the sub-optimality that might be induced by using an a priori structure-frozen explicit feedback instead of the ideal implicit moving-horizon state-dependent optimal control (also called Model Predictive Control in the control community). 

Although the methodology proposed in this paper is intended to have a general scope (provided that a problem-dependent parameterized control feedback is designed following the same kind of methodologies described in this paper), it is illustrated using the example of mixed therapy of cancer involving three combined drugs, namely: chemotherapy, vaccine and immunotherapy drugs \cite{alamir_robust_cancer:2014}. The population model is a $6$-th dimensional model involving 24 parameters which enables to show the scalability of the proposed method as well as the relevance of the underlying global sensitivity analysis framework. 

This paper is organized as follows: First the specific use-case of mixed therapy of cancer is presented in Section \ref{sec-example} since it makes easier to follow the general presentation of the methodology proposed in Section \ref{sec-methodology} by referring to concrete instantiations of each of the general concepts and definitions. Section \ref{sec-results} shows the application of the methodology to the supporting example of combined therapy of cancer before a general conclusion and some hints for further investigation are given in Section \ref{sec-conc}.

\section{The supporting example: combined chemotherapy/vaccine/immunotherapy of cancer}\label{sec-example}
In this section, the supporting example that is used in the validation of the proposed framework is presented. First the mathematical equations are described in Section \ref{sec-equation} before the control problem is stated in section \ref{sec-control-problem}.
\subsection{The population model's equations}\label{sec-equation}
In this section, the mathematical model proposed in \cite{DePillis06,DePillis:01} which governs the dynamics of the population of cells in presence of tumor and under the combined action of chemotherapy, vaccine and immunotherapy is briefly described. This model serves as a supporting example illustrating the methodology in the remainder of this paper. The model is based on published statements and quite nicely fits clinical validation \cite{Rosenberg1986,Diefenbach2001,Dudley2002}. The underlying assumptions that support this model are described in details in \cite{Chareyron:09}\footnote{See table 1 page 445.}. The model involves the following cell populations:
\begin{itemize}
\item[$\checkmark$] $T$: the population of tumor cells. 
\item[$\checkmark$] $N$: the total NK cells population which are part of the innate immune system in so far as they exist even in the absence of tumor. 
\item[$\checkmark$] $L$: the total CD8$^+$T cells population which are active cells that are tumor specific which carry the immune response when tumor cells are present in the body.
\item[$\checkmark$] $C$: the population of white blood cells (circulating lymphocytes). The size of this population can be viewed as a relevant indicator of the patient's health. 
\item[$\checkmark$] $M$, $I$: Chemotherapy and Immunotherapy drugs concentrations in the blood stream.
\end{itemize}  
On the other hand, the three control inputs are:
\begin{itemize}
\item[$\checkmark$] $v_M$: the chemotherapy drug injection rate;
\item[$\checkmark$] $v_I$: the immunotherapy drug injection rate and
\item[$\checkmark$] $v_L$: the vaccin injection rate
\end{itemize} 
Using these states and control variables, the dynamic model governing the evolution of the population sizes is given by the following set of Ordinary Differential Equations (ODEs):
{\footnotesize
\begin{subequations}
\begin{align}
\dfrac{dT}{dt}&=\Bigl[a(1-bT)-cN-D-K_T(1-e^{-M})\Bigr]T \label{dTdt}\\
\dfrac{dN}{dt}&=eC+\Bigl[-f+g\frac{T^2}{h+T^2}-pT-K_N(1-e^{-M})\Bigr]N\label{dNdt}\\
\dfrac{dL}{dt}&=-mL+j\frac{D^2T^2}{k+D^2T^2}L-qLT+(r_1N+r_2C)T-\nonumber \\
&-uNL^2-k_L(1-e^{-M})L+\frac{p_I I}{g_I+I}L+v_L(t)\label{dLdt}\\
\dfrac{dC}{dt}&=\alpha-\beta C-k_C(1-e^{-M})C \label{dCdt}\\
\dfrac{dM}{dt}&=-\gamma M + v_M(t)\label{dMdt}\\
\dfrac{dI}{dt}&=-\mu_I I + v_I(t)\label{dIdt}\\
\end{align}
\end{subequations}
}
\ \\ 
To summarize, the above model involves six measured state variables, namely $T$, $N$, $L$, $C$, $M$ and $I$, three manipulated variables $v_M$, $v_L$ and $v_I$ and 24 model's parameters (all the remaining lower-cased symbols, namely: $a, b, c, \dots$). Note however that the term $D$ involved in \eqref{dTdt} is a composite term defined by:
\begin{equation}
D:=\left[\dfrac{(L/T)^\ell}{s+(L/T)^\ell}\right]d \label{eqD}
\end{equation} 

In the remainder of this paper, the above set of ODEs will be referred to using the following compact form:
\begin{equation}
\dot x=f(x,u,p_\text{model}) \label{shorteq}	
\end{equation}
where $x:=(T,N,L,C,M,I)\in \mathbb R^{n_x}$ ($n_x=6$) is the state vector, $u=(v_M, v_I, v_L)\in \mathbb R^{n_u}$ ($n_u=3$) is the manipulated (control) input while $p_\text{model}\in \mathbb R^{n_p}$ ($n_p=24$) gathers the parameters of the model that are supposed to be imperfectly known. Moreover, it is assumed that the dispersion of the parameters is characterized by a {\bf known} probability distribution which enables to draw relevant set of samples when necessary.  \\ \ \\
Finally, it is assumed that the drugs injection rates are bounded by appropriate upper bounds, namely
\begin{equation}
v_\sigma \in [0, \bar v_\sigma]\quad ;\quad \sigma \in \{M,I,L\} \label{saturv}
\end{equation}  
\subsection{The therapy as a control problem}\label{sec-control-problem}
The paradigm of drug dosage and scheduling in cancer therapy is determined by two antagonistic requirements, namely: the contraction of the tumor at the end of the treatment duration, say $T_{th}$, with a sufficient contraction factor $\gamma_c\ll 1$ (constraint \eqref{TcontractionCond} hereafter) and the need to maintain the population of circulating lymphocytes $C$ above some minimum value $C_\text{min}>0$ during the whole treatment duration (constraint \eqref{healthCond} hereafter). 
\begin{figure}[H]
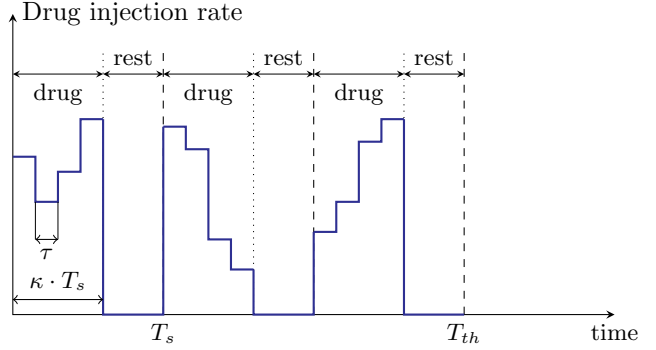

\begin{center}
\tikz{
\def\Lx{8}
\def\Ly{4}
\def\ytop{3.5}
\def\Tsub{2}
\def\Ttrait{1.2}
\def\yrest{3.2}
\def\dt{0.3}
\draw[<->,>=stealth] (\Lx,0)node[below]{\footnotesize time} -- (0,0) -- (0,\Ly) node[right]{Drug injection rate};
\draw[dashed] (\Tsub,0) node[below]{\footnotesize $T_s$}-- (\Tsub,\ytop);
\draw[dashed] (2*\Tsub,0) -- (2*\Tsub,\ytop);
\draw[dashed] (3*\Tsub,0)node[below]{\footnotesize $T_{th}$} -- (3*\Tsub,\ytop) ;
\draw[dotted] (\Ttrait,0) -- (\Ttrait,\ytop);
\draw[dotted] (1*\Tsub+\Ttrait,0) -- (\Tsub+\Ttrait,\ytop);
\draw[dotted] (2*\Tsub+\Ttrait,0) -- (2*\Tsub+\Ttrait,\ytop);
\draw[very thin,<->,>=stealth] (\Ttrait,\yrest) -- node[midway,above]{\footnotesize rest} (\Tsub,\yrest);
\draw[very thin,<->,>=stealth] (\Ttrait+\Tsub,\yrest) -- node[midway,above]{\footnotesize rest} (2*\Tsub,\yrest);
\draw[very thin,<->,>=stealth] (\Ttrait+2*\Tsub,\yrest) -- node[midway,above]{\footnotesize rest} (3*\Tsub,\yrest);
\draw[very thin,<->,>=stealth] (0,\yrest) -- node[midway,below]{\footnotesize drug} (\Ttrait,\yrest);
\draw[very thin,<->,>=stealth] (\Tsub,\yrest) -- node[midway,below]{\footnotesize drug} (\Ttrait+\Tsub,\yrest);
\draw[very thin,<->,>=stealth] (2*\Tsub,\yrest) -- node[midway,below]{\footnotesize drug} (\Ttrait+2*\Tsub,\yrest);
\draw[thick,Blue] (0,2.1) -- (\dt,2.1) -- (\dt,1.5) -- (2*\dt,1.5) -- (2*\dt,1.9) -- (3*\dt,1.9) -- (3*\dt,2.6) -- (4*\dt,2.6);
\draw[thick,Blue] (4*\dt,2.6) -- (4*\dt,0) -- (0+\Tsub,0) -- (0+\Tsub,2.5);
\draw[thick,Blue] (0+\Tsub,2.5) -- (\dt+\Tsub,2.5) -- (\dt+\Tsub,2.2) -- (2*\dt+\Tsub,2.2) -- (2*\dt+\Tsub,1) -- (3*\dt+\Tsub,1) -- (3*\dt+\Tsub,0.6) -- (4*\dt+\Tsub,0.6);
\draw[thick,Blue] (4*\dt+\Tsub,0.6) -- (4*\dt+\Tsub,0) -- (2*\Tsub,0) -- (2*\Tsub,1.1);
\draw[thick,Blue] (0+2*\Tsub,1.1) -- (\dt+2*\Tsub,1.1) -- (\dt+2*\Tsub,1.5) -- (2*\dt+2*\Tsub,1.5) -- (2*\dt+2*\Tsub,2.3) -- (3*\dt+2*\Tsub,2.3) -- (3*\dt+2*\Tsub,2.6) -- (4*\dt+2*\Tsub,2.6);
\draw[thick,Blue] (4*\dt+2*\Tsub,2.6) -- (4*\dt+2*\Tsub,0) -- (3*\Tsub,0);
\draw[solid, very thin] (\dt,1.5) -- (\dt,1);
\draw[solid, very thin] (\dt+\dt,1.5) -- (\dt+\dt,1);
\draw[<->] (\dt,1) -- node[midway,below]{\footnotesize $\tau$} (2*\dt,1);
\draw[<->,thin] (0,0.2) -- node[above]{\footnotesize $\kappa\cdot T_s$}(\Ttrait,0.2);
}
\end{center} 
\caption{\color{Blue} Temporal structure of the therapy. Example of a treatment period consisting of $N_T=3$ sub-periods of a duty cycle $\kappa\in (0,1)$.  Note that the $y$-axis represents the injection rates of the different combined drugs and as such, the plot schematically represents a vector evolution and not a single variable. The periods of rest are periods during which no treatment of any kind is applied. The ratio between the treatment periods and the rest periods is the duty cycle $\kappa$. The values of the injection rates are defined by a state feedback whose parameters are to be optimized so that some outcome of the therapy can be certified despite of the high uncertainties on the model's parameters.}  \label{fig_protocol} 
\end{figure}
On the other hand, for obvious reasons, a temporal constraint has to be imposed on the drug delivery schedule which can take place only during specific intervals of time that are separated by rest periods during which drug cannot be delivered. This constraint is depicted on Figure \ref{fig_protocol}. More precisely, this Figure shows a therapy schedule lasting $T_{th}$ days which involves three basic periods of duration $T_s$, each of which is decomposed into a treatment sub period of duration $\kappa T_s$ followed by a rest sub-period of duration $(1-\kappa)T_s$ where $\kappa\in (0,1)$. Note also that during the drug delivery sub-intervals, the injection rate is forced to be piece-wise constant with an elementary sampling period of $\tau$ (This might be equal to 3h, 6h, 12h or such) since continuous adaptation is not compatible with real-life mode of delivery and measurement accessibility.

Based on the above notation, the $T_{th}$-lasting therapy is considered to be successful, for a given level of contraction $\gamma_c$, if the following conditions hold true:
\begin{subequations}
\begin{align}
&\text{(Tumor contraction)}\qquad  &T(T_{th})\le \gamma_c T(0) \label{TcontractionCond} \\
&\text{(Health condition)}\qquad &\min_{t\le T_{th}}C(t)\ge C_\text{min} \label{healthCond}
\end{align} 
\label{conditions}
\end{subequations}
Note however that the quantities involved in the l.h.s of \eqref{conditions}, namely $T(T_{th})$ and $C(t)$ depend on the following:
\begin{itemize}
\item[$\checkmark$] The initial state $x(0)=x_0$ at the very beginning of the therapy;
\item[$\checkmark$]The specific patient being treated which is represented by a specific realization of the vector of parameters $p_\text{model}\in \mathbb R^{n_p}$ and\\
\item[$\checkmark$] The piece-wise constant feedback strategy that is used during the drug-delivery sub-periods, namely:
\begin{equation}
\forall t\in [k\tau, (k+1)\tau)\qquad u(t) = K(x(k\tau), p_{ctr})	
\end{equation}
where $[k\tau, (k+1)\tau)$ is an elementary interval that lies in a drug delivery sub-period (see Figure \ref{fig_protocol}). Note that during this elementary interval of duration $\tau$, the injection rate is constant and depends on the value of the state vector at the beginning of this interval, namely $x(k\tau)$. Note also that the feedback strategy $K$ is defined by a vector of $n_c$ control design parameters which is denoted by $p_{ctr}\in \mathbb R^{n_c}$. The specific instantiation of this vector in the specific case of the supporting example of mixed therapy is explained later on (Section \ref{feedbackdesign}).
\end{itemize}
Based on the above, it comes out that the constraints \eqref{conditions} can be rewritten in a form that involves the underlying arguments in a more explicit way, namely:
\begin{subequations}
\begin{align}
&\text{(Contraction)}\   &C_1(x_0, p_\text{model}, p_\text{ctr}, \gamma_c)\le 0\label{TcontractionCondA} \\
&\text{(Health)}\quad &C_2(x_0, p_\text{model}, p_\text{ctr})\ge C_\text{min} \label{healthCondA}
\end{align} 
\label{conditionsA}
\end{subequations}
for a straightforward appropriate definitions of the maps $C_1$ and $C_2$ that reflect the constraints \eqref{conditions}.

The design of the feedback strategy as well as the choice of the vector of control parameters $p_{ctr}$ it involves is a part of the therapy degrees of freedom to which one should add the choice of the basic interval duration $T_s$ and the duty ratio $\kappa\in (0,1)$ which defines the effective treatment part of this basic interval, namely $\kappa T_s$. The concatenation of these degrees of freedom results in the therapy parameter vector defined by
\begin{equation}
\text{\footnotesize(Therapy's parameters)}\qquad p_\text{therapy}:= \begin{bmatrix}
p_{ctr}\cr T_s\cr \kappa	
\end{bmatrix} \label{defdeptherapy}
\end{equation} 
which assumes that the total duration $T_{th}$ of the therapy is fixed a priori\footnote{Note that nothing prevents from considering this duration as a decision variable in the proposed framework.}. 

\section{Description of the methodology}\label{sec-methodology}
The methodology proposed in this paper involves the following steps:\\ \ \\
(1) {\bf Design a parameterized feedback strategy}: $$K(x,p_\text{ctr})\quad p_\text{ctr}\in \mathcal P_\text{ctr}$$ based on simple principles and on the understanding of the path linking the drug delivery to the evolution of the key variables. An example of such design is given hereafter (Section \ref{feedbackdesign}). The subset $\mathcal P_\text{ctr}\subset \mathbb R^{n_c}$ is the set of possible values of the control parameters. Note that this design step is strongly problem-dependent. Any existing state feedback (including those that do not explicitly handle the uncertain nature of the model's parameters) might be used provided that it involves some tunable set of parameters gathered in the vector $p_\text{ctr}$. \\ \ \\
(2) {\bf Generate a set $\mathbb P_\text{model}$ of $n_R$ realizations} of the vector of parameters of the model according to the known (or presumed) probability distribution, namely:
\begin{align}
&\text{card}(\mathbb P_\text{model})=n_R\\
&\mathbb P_\text{model}:=\bigl\{p^{(1)}_\text{model}, \dots, p^{(n_R)}_\text{model}\bigr\}\subset \mathbb R^{n_p}	
\end{align}
where for all $i$, $p^{(i)}_\text{model}\in \mathbb R^{n_p}$ stands for a specific realization of the uncertain model's vector of parameters. It is assumed that a nominal value $p_\text{model}^\text{nom}$ of the parameter vector is chosen to be used in the design of the nominal feedback strategy $K(x,p_\text{ctr})$ that is extensively used in the sequel. \\ \ \\
(3) {\bf Generate a set $\mathbb X$ of $n_R$ initial conditions} uniformly distributed inside a hyper-box $[\underline x, \bar x]\subset \mathbb R^{n_x}$, namely
\begin{align}
&\text{card}(\mathbb X)=n_R\\
&\mathbb X:=\bigl\{x^{(1)}_0, \dots, x^{(n_R)}_0\bigr\}\subset \mathbb R^{n_x}	
\end{align}
Note that $\underline x\in \mathbb R^{n_x} $ and $\bar x\in \mathbb R^{n_x} $ stand for the lower and upper bounds of the possible values of the state vector. \\ \ \\
(4) {\bf Generate a set $\mathbb P_\text{ctr}$ of $n_R$ values of the control} parameter vector by random sampling inside the admissible set $\mathcal P_\text{ctr}$, namely:
\begin{align}
&\text{card}(\mathbb P_\text{ctr})=n_R\\
&\mathbb P_\text{ctr}:=\bigl\{p^{(1)}_\text{ctr}, \dots, p^{(n_R)}_\text{ctr}\bigr\}\subset \mathbb R^{n_c}	
\end{align}
(5) {\bf Simulate the cloud of closed-loop trajectories}. Having the set of triplets:
\begin{equation}
\mathcal Z:=\Bigl\{z^{(i)}:=(x_0^{(i)}, p^{(i)}_\text{model}, p^{(i)}_\text{ctr})\Bigr\}_{i=1}^{n_R}\label{defdeZ}
\end{equation} 
The corresponding set of $n_R$ closed-loop trajectories can be obtained by simulating the closed-loop dynamics that is defined for all $t\in [k\tau, (k+1)\tau)$ by [see \eqref{shorteq}]:
\begin{equation*}
\dot x(t) = f\Bigl(x(t),K(x(k\tau),p^{(i)}_\text{ctr}), p^{(i)}_\text{model}\Bigr)\quad x(0)=x_0^{(i)}
\end{equation*} 
representing the predicted evolutions of the population sizes under the time-sampled control strategy $K(\cdot, p^{(i)}_\text{ctr})$, for a specific set of model's parameters $p^{(i)}_\text{model}$ and starting from the initial state $x^{(i)}_0$.

In the remainder of this presentation, the vector $z^{(i)}:=(x_0^{(i)}, p^{(i)}_\text{model}, p^{(i)}_\text{ctr})$ is referred to as the \texttt{features vector} associated to the $i$-th sample of the underlying learning cloud.\ \\ \ \\
(6) {\bf Analyze the simulated trajectories}. In this step, each closed-loop trajectory associated to the \texttt{feature vector} $z^{(i)}$ is analyzed in order to derive the following \texttt{labels} to be later used in building ML predictors:
\begin{itemize}
\item[$\checkmark$] The contraction boolean label: 
\begin{equation}
y_T^{(i)}:=\Bigl[C_1(z^{(i)}, \gamma_c)\le 0\Bigr]\in \{\texttt{True}, \texttt{False}\}\label{yC}
\end{equation} 
 where $C_1\le 0$ is the contraction condition \eqref{TcontractionCondA}. \\
\item[$\checkmark$] The health boolean label:  
\begin{equation}
y_H^{(i)}:=\Bigl[\dfrac{C_2(z^{(i)})-C_\text{min}}{C_\text{min}}\ge \rho\Bigr]\in \{\texttt{True}, \texttt{False}\}\label{yH}
\end{equation} 
which is a strongest tightened version of the health constraint \eqref{healthCondA} [that would be obtained for $\rho=0$]. 
\item[$\checkmark$] The amount of drugs delivered during the therapy:
\begin{equation}
y_j^{(i)}:= \dfrac{1}{\bar v_j}\int_0^{T_{th}}v_j^{(i)}(\tau)d\tau\ ;\quad j\in \{M, I, L\} \label{vj}
\end{equation} 
where $\bar v_j$ denotes the maximum allowed injection rate of drug $j$. 
\end{itemize} 
Once these five labels are computed for all the features vectors $z^{(i)}$, $i=1,\dots,n_R$ included in $\mathcal Z$, one can concatenate them to get five vectors of labels associated to the matrix of features $\mathcal Z$ defined in \eqref{defdeZ}, namely:
\begin{equation}
\mathcal Y_\sigma:= \Bigl\{y_\sigma^{(i)}\Bigr\}_{i=1}^{n_R}\quad;\quad \sigma\in \{T, H, M, I, L\}
\end{equation} 
(7) {\bf Fit a prediction model $\mathbf{F_\sigma}$} for each value of $\sigma$. Indeed, for each $\sigma\in \{T, H, M, I, L\}$, a Machine Learning problem can be defined by the following learning data:
\begin{equation}
\mathcal D_\sigma := \Bigl\{\mathcal Z, \mathcal Y_\sigma\Bigr\}
\end{equation} 
More precisely a ML problem is the problem of finding a prediction map $F_\sigma$ defined on the space of features and taking values in the space of labels, such that the following approximation holds with a sufficiently high precision:
\begin{equation}
\forall (z,y_\sigma)\in \mathcal Z\times \mathcal Y_\sigma\qquad F_\sigma(z)\approx y_\sigma
\end{equation} 
When $y_\sigma\in \{\texttt{true}, \texttt{false}\}$ ($\sigma\in \{T, H\}$), the map $F_\sigma$ is called a binary classifier and the ML problem is a binary classification problem. When $y_\sigma\in \mathbb R$ ($\sigma\in \{M, I, L\}$), the map $F_\sigma$ is called a regressor and the ML associated problem is called a regression problem.\\ \ \\
(8) {\bf Perform a global sensitivity analysis} on the so fitted models for the tumor contraction map $F_T$ and for the health monitoring map $F_H$ in order to rank the order of the \texttt{features} (component of $z=(x, p_\text{model}, p_\text{ctr})$)  that really matter for the quality of the associated prediction. Note that this global sensitivity analysis comes as a side product when the fit is performed using specific available models such as the \texttt{scikitlearn}'s \texttt{RandomForestClassifier} \cite{scikit-learn} that is used hereafter.  \\ \ \\
Once a reduced number of parameters are chosen, the models are refitted using the so-chosen \texttt{features} in order to validate this dimension reduction step. \ \\ \ \\
(9) Use the reduced size selected \texttt{features} to derive relevant information via off-line optimization that might be presented on the reduced state subspaces (see Section \ref{sec-results} for an instantiation of this step for the combined therapy's illustrative example.  
\section{Application}\label{sec-results}
In this section, the roadmap described in Section \ref{sec-methodology} is applied to the specific multi-therapy of cancer problem defined in Section \ref{sec-example}.
\subsection{Design of explicit parameterized feedback strategy} \label{feedbackdesign}
The feedback design is based on the following rules:
\begin{enumerate}
\item Chemotherapy injection must stop if the Tumor is lower that some lower bound, denoted by $T_{stop}$~t:
\begin{equation}
T<T_{stop}\quad \Rightarrow \quad v_M = 0\label{Tlow}
\end{equation} 
Note that $T_{stop}$ is the first parameter of the control strategy (one of the components of the vector of parameters $p_\text{ctr}$).\\
\item All drug injection must stop if the rate of decrease of the tumor is \textit{Sufficient}. This condition writes:
\begin{equation}
\dot{T}\le -rT \quad \Rightarrow \quad v_M=v_I=v_L=0\label{goodincrease}
\end{equation} 
where $r>0$ is another component of the vector of control parameters $p_\text{ctr}$. The rationale behind these first two conditions is to reduce the use of drugs and avoid injecting chemotherapy when the tumor is too small or when the natural drug-free dynamics corresponds to a sufficient decrease rate of the tumor. Obviously, whether a value $r$ corresponds to a sufficient decrease cannot be decided intuitively, that is the reason why $r$ is viewed as degree of freedom to be determined via the learning process and the optimization that is made possible by the availability of the fitted prediction models. \\
\item Since a non zero concentration $M$ of chemotherapy drug might induce a further decrease in the circulating lymphocytes $C$, it is mandatory that this concentration approches $0$ when $C$ approaches the lower bound $C_{min}$ or a tightened version $\beta_CC_{min}$ ($\beta_C>1$) of it for security reasons in order to take into account the bad knowledge of the parameters. This suggest the following definition of the chemotherapy injection:\\ \ \\
\tikz{
\node[fill=Gray!15, rounded corners] at(0,0)(vMdef){
\begin{minipage}{0.42\textwidth}
\begin{equation}
v_M = \left\{\begin{array}{l}
0 $\quad$ \text{if ($T\le T_{stop}$) or ($\dot T\le -rT$)}	\\
\min\Bigl\{\bar v_M, \max\bigl\{0,\mu_C(C-\beta_CC_{min})\bigr\}\Bigr\}
\end{array}
\right. \label{betaCmuC}
\end{equation}
\end{minipage} 
};
\node[below=0.1mm] at(vMdef.north){\footnotesize \bf \color{Blue} Chemotherapy};
}
\ \\
where $\mu_C>0$ and $\beta_C>1$ are components of the vector of control parameters $p_\text{ctr}$. Indeed, when $C\le \beta_CC_{min}$, the definition induces $v_M=0$, otherwise chemotherapy injection is allowed with a rate that depends on $\mu_C>0$ while keeps saturated at the allowable maximum injection rate $\bar v_M$. This completely defines the feedback strategy regarding the chemotherapy injection rate. The remaining following rule completes the definition of the feedback strategy for vaccine and immunotherapy injection rates. \\
\item The dynamics of population size described above leads to the following graph linking the immunotherapy drug and vaccine injection rates $v_I$ and $v_L$ to the decrease of the tumor size $T$:\\ \ \\
\tikz{
\node at(0,0)(vI){$v_I$};
\node at(0,-0.8)(vL){$v_L$};
\node at(1.6,0)(I){$I$};
\node at(3.2,0)(L){$L$};
\node at(4.8,0)(D){$D$};
\node at(6.4,0)(T){$T$};
\draw[->] (vI) --node[below]{\tiny \eqref{dIdt}} (I);
\draw[->] (vI) --node[above]{\tiny increase} (I);
\draw[->] (I) --node[above]{\tiny increase} (L);
\draw[->] (I) --node[below]{\tiny \eqref{dLdt}} (L);
\draw[->] (L) --node[above]{\tiny increase} (D);
\draw[->] (L) --node[below]{\tiny  \eqref{eqD}} (D);
\draw[->] (D) --node[above]{\tiny decrease} (T);
\draw[->] (vL.east) -| node[pos=0.2, above]{\tiny increase}(L.south);
\draw[->] (vL.east) -| node[pos=0.2, below]{\tiny \eqref{dLdt}}(L.south);
}
which clearly shows that the corresponding decrease in the tumor arises only through the term $D$. But the definition \eqref{eqD} of $D$ implies that:
\begin{equation*}	
D\le d 
\end{equation*}
meaning that when $D$ approaches $d$, there is no noticeable benefit from further injecting immunotherapy or vaccine drugs. This suggests the following definition of these drugs' injection rate:\\ \ \\
\tikz{
\node[fill=Gray!15, rounded corners] at(0,0)(vMdef){
\begin{minipage}{0.42\textwidth}
\begin{align}
&\forall \sigma\in \{I, L\}, \nonumber \\
&v_\sigma = \left\{\begin{array}{ll}
0& \text{if}\ \dot T\le -rT \\	
&\min\Biggl\{1, \Bigl[\dfrac{D_{max}-D}{D_{max}}\Bigr]\Bigl[\dfrac{T}{T_{max}}\Bigr]\Biggr\} \cdot \bar v_\sigma
\end{array}\right.\label{stratimmuno}
\end{align} 
where $D_{max}=c_d\cdot d^{nom}$ with $c_d>1$.
\end{minipage} 
};
\node[below=0.05mm] at(vMdef.north){\footnotesize \bf \color{Blue} Immunotherapy drug and vaccine};
}
\ \\
where the use of the control parameter $c_d>1$ induce a secure version of $D_{max}=d$ that acknowledges that the model's parameter $d$ might differ from the nominal value $d^{nom}$. $T_{max}$ is some reference value for high level of tumor sizes beyond which it can be considered that successful therapy is no more a realistic option. \\ \ \\ By so defining $v_\sigma$, the constraint \eqref{saturv} is always satisfied on one hand. On the other hand, when $D$ approaches its maximum value or when the tumor is too small, then the immunotherapy and the vaccin injection rates are reduced correspondingly. \\ \ \\
This adds a new component\footnote{$T_{max}$ if a priori fixed to $10^{8}$.} to the parameter vector $p_\text{ctr}$, namely $c_d$. Based on the discussion above, the vector of control parameter $p_\text{ctr}$ is given by:
\begin{equation}
p_\text{ctr}:= \begin{bmatrix}
T_{stop}&r&\mu_C&\beta_C&c_d
 \end{bmatrix} \label{defdepctr}
\end{equation}   
which is to be used in \eqref{defdeptherapy} in order to get the full definition of the degrees of freedom of the whole therapy.
\end{enumerate}
\subsection{Generate the learning dataset} \label{SecgenerateLData}
\subsubsection{Generation of the set $\mathbb P_\text{model}$ (model realizations)}
For each level of uncertainties 
\begin{equation}
\zeta\in \mathcal S_p:=\Bigl\{0\%, 10\%, 20\%, 30\%, 40\%, 50\%, 80\%\Bigr\}
\end{equation} 
A set $\mathbb P_\zeta$ composed of 10000 samples are uniformly drawn inside the interval $[(1-\zeta)p^\text{nom}, (1+\zeta)p^\text{nom}]$, then the set $\mathbb P_\text{model}$ is defined as follows:
\begin{equation}
\mathbb P_\text{model}:= \bigcup_{\zeta\in \mathcal S_p} \mathbb P_\zeta
\end{equation} 
Note that by so doing, one gets 
\begin{equation}
n_R:= \text{card}\bigl(\mathbb P_\text{model}\bigr)=7\times 10^4
\end{equation} 
The nominal values of the parameters are are given in \cite{DePillis06} (see table 1). Moreover they are used in the programs that are made accessible for download from the \texttt{Github} site of the author\footnote{https://zenodo.org/badge/latestdoi/492703016}. 
\subsubsection{Generation of the set $\mathbb X$ (initial states)}
The set of initial state $\mathbb X$ is obtained by uniform sampling $n_R$ values of the initial state in the hypercube $[\underline x, \bar x]$ where:
\begin{align*}
\underline x&:=\begin{bmatrix}
10^5& 10^{-3}& 10^{-3}&1.05\times C_{min}&10^{-3}&10^{-3}
\end{bmatrix} \\
\bar x&:=\begin{bmatrix}
10^9& 10^{3}& 10^{8}&10^{11.1}&10^{-3}&10^{-3}
\end{bmatrix}
\end{align*} 
Note that the lower and upper values on the initial concentration of the drugs is supposed to be almost equal to $0$ since the therapy begins at instant $0$ when the patient is admitted at the hospital. On the other hand, the lower value of the circulating lymphocytes size is also equal to 5\% above its minimal value which is compatible with the previous assumption. The lower value ($10^5$) of the initial tumor is used because smaller values correspond to favorable initial conditions that can be removed as they correspond to successful therapy for almost any sampled model's parameters and any chosen control parameters. 
\subsubsection{Generation of the set $\mathbb P_\text{ctr}$ (control parameters)}
The set $\mathbb P_\text{ctr}$ is obtained by sampling $n_R$ values of the parameters uniformly inside the hypercube defined by the minimum and the maximum values shown on Table \ref{minmaxvalpar} except for the the basic period $T_s$ that is taken randomly to be one of the three values shown in Table \ref{minmaxvalpar}.

\begin{table}[H]
\begin{center}
\begin{tabular}{lccr} \toprule
    {\bf Parameter} & {\bf min-value} & {\bf max-value} & {\bf Equation} \\ \midrule
$T_\text{stop}$&  $10^1$&  $10^3$&  \eqref{Tlow}\\
 $r$&  $10^{-1}$&  $10^1$&  \eqref{goodincrease}\\
 $\beta_C$&  $10^{1}$&  $10^2$&  \eqref{betaCmuC}\\
 $\mu_C$&  $10^{-1}$&  $1$&  \eqref{betaCmuC}\\
 $c_d$&  $0.8$&  $1.5$&  \eqref{stratimmuno}\\
 $\kappa$&  $0.2$&  $0.9$&  (Figure \ref{fig_protocol})\\
 $T_s$ &  $\in\{0.5, 1, 2\}$&  &(Figure \ref{fig_protocol})\\\bottomrule \\
\end{tabular}
\end{center} 
\caption{\color{Blue} Minimum and maximum values of the control parameters used in the generation of the learning dataset.}
\label{minmaxvalpar}
\end{table}
\subsubsection{Fitting prediction models}\label{fittingpredictionmodel}
\begin{figure*}
\begin{center}
\includegraphics[width=0.53\textwidth]{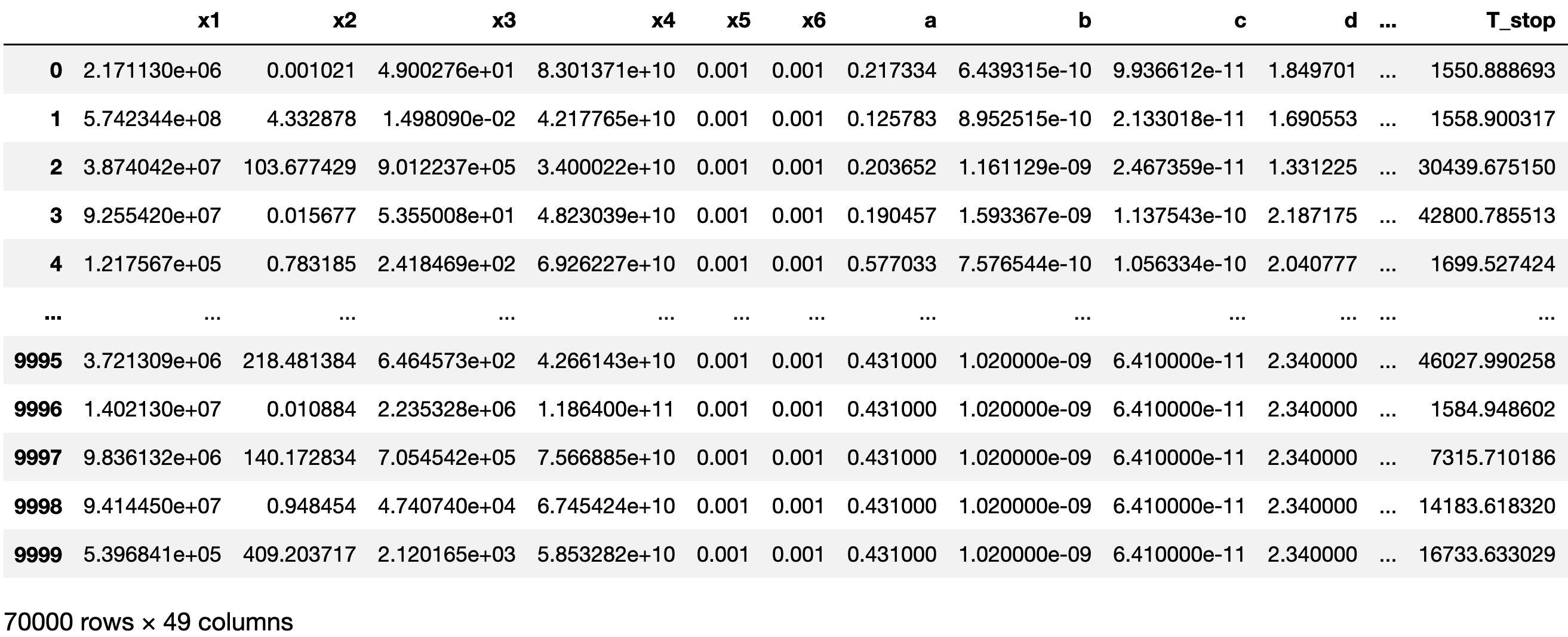} \includegraphics[width=0.43\textwidth]{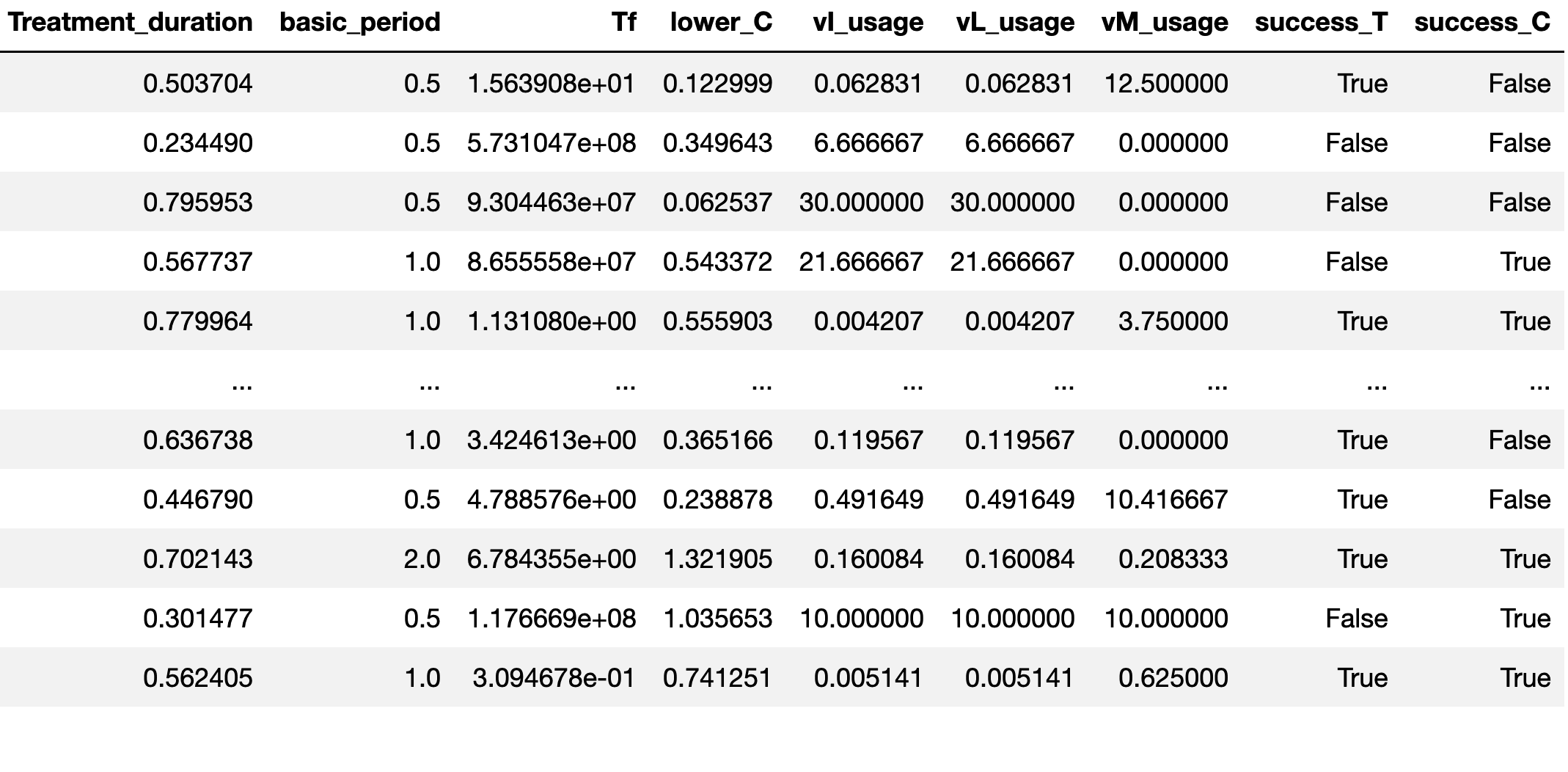} 
\end{center} 	
\caption{\color{Blue} View of the dataframe representing the learning data involving 70000 samples of initial state, model's parameters and control parameters. The last two columns represents the boolean labels regarding the fulfillment of the tumor contraction condition and the condition of the health indicator being safely above the threshold. The previous last columns show the normalized quantities of the three drugs [see \eqref{vj}] being used over the therapy duration. $T_f$ stands for the final values of the tumor.}\label{fig-learning-data}
\end{figure*}
\begin{figure*}
\begin{center}
\includegraphics[width=0.4\textwidth]{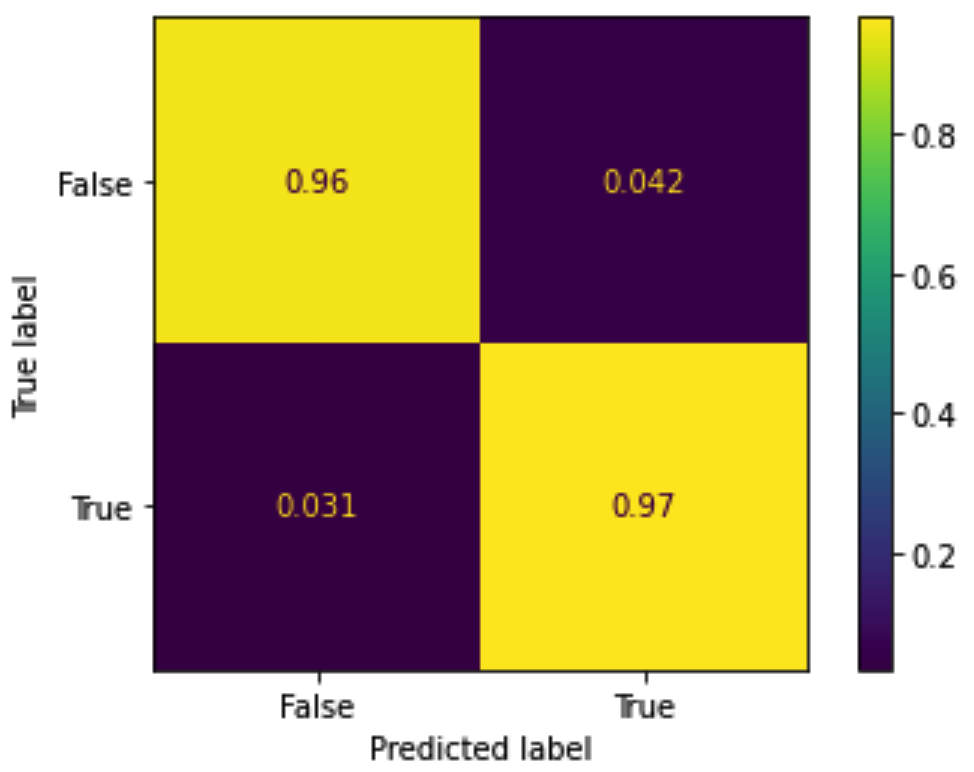} \includegraphics[width=0.4\textwidth]{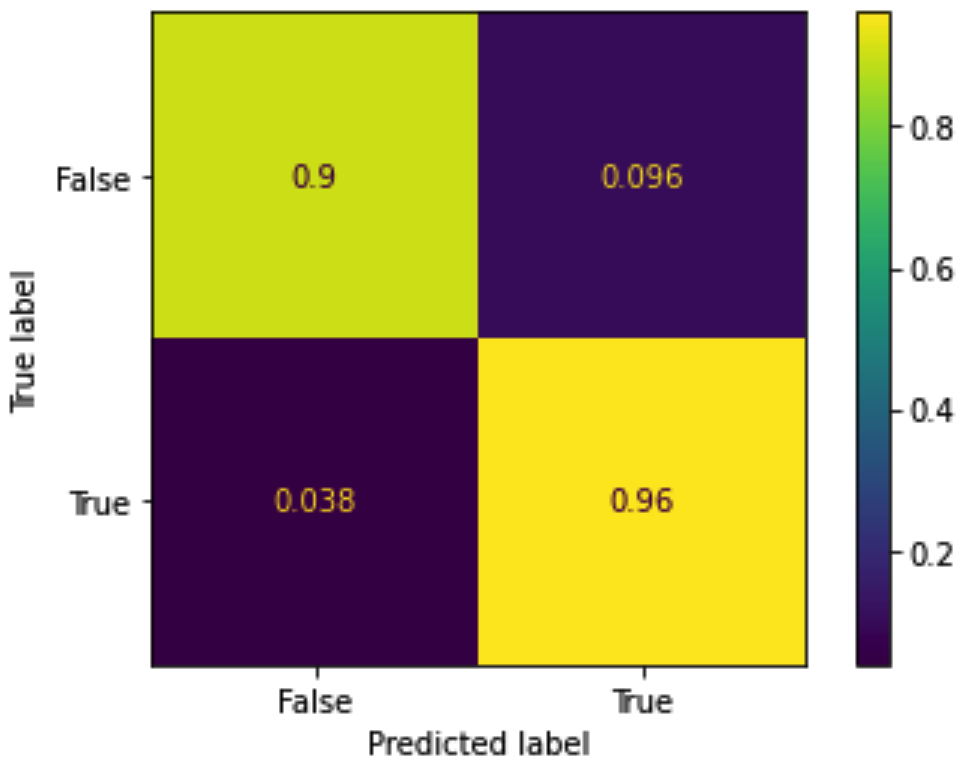} 
\end{center} 	
\caption{\color{Blue} Prediction results on the 21000 scenarios (30\% of 70000) included in the test data using all the \texttt{features}. {\bf (Left)} Prediction regarding the tumor contraction. {\bf (Right)} Prediction of the health condition satisfaction. The \texttt{RandomForestClassifier} has been trained using 100 estimators while limiting the maximum number of leaves to 2000.}\label{fig-prediction-all-binary}
\end{figure*}
\begin{figure*}
\begin{center}
\includegraphics[width=0.4\textwidth]{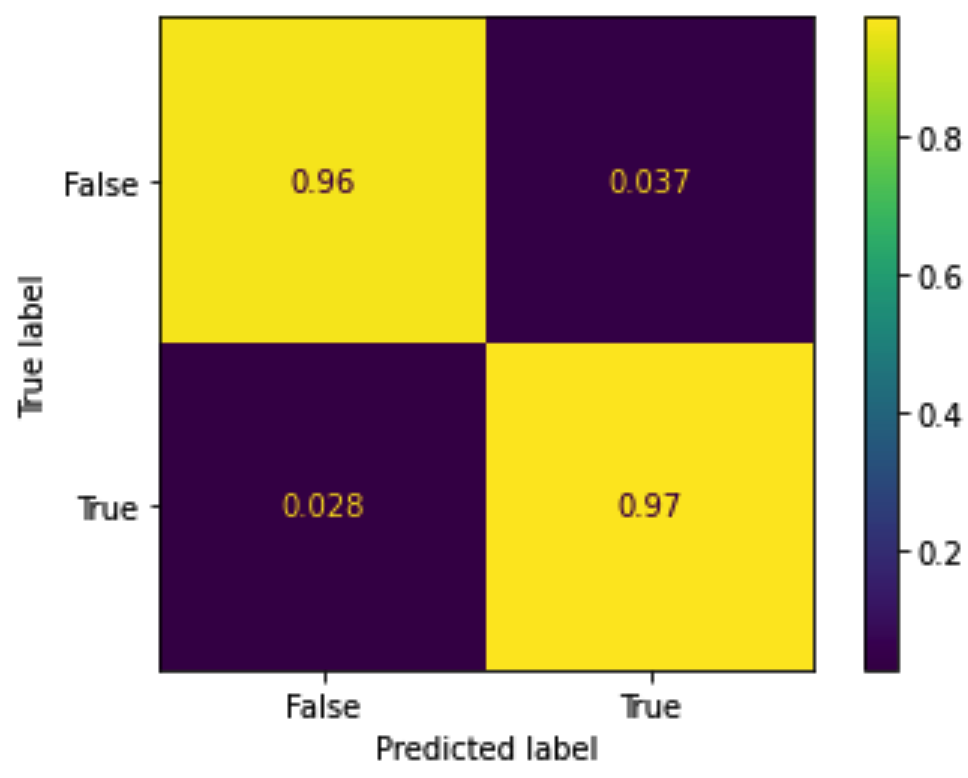} \includegraphics[width=0.4\textwidth]{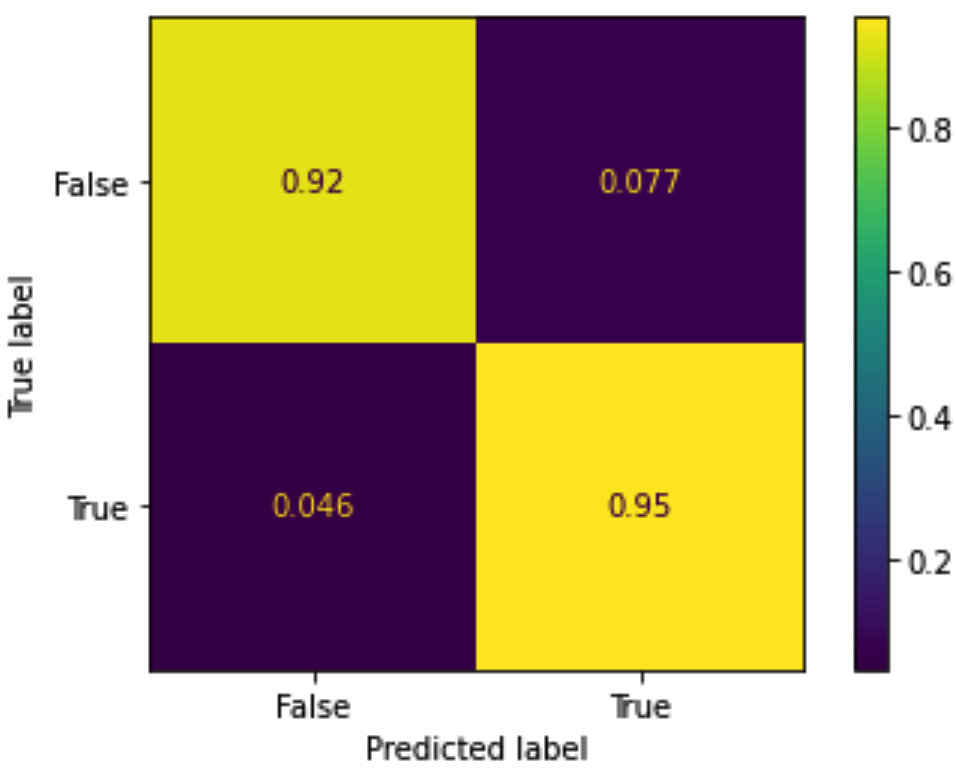} 
\end{center} 	
\caption{\color{Blue} Prediction results on the 21000 scenarios (30\% of 70000) included in the test data using only the important \texttt{features} shown in Table \ref{tableimportantfeatures}. {\bf (Left)} Prediction regarding the tumor contraction. {\bf (Right)} Prediction of the health condition satisfaction. }\label{fig-prediction-selected-binary}
\end{figure*}

As explained in item (5) of Section \ref{sec-methodology}, a set of $n_R=70000$ closed-loop simulation are conducted to build the so called learning data. The data frame representing these simulations is shown in Figure \ref{fig-learning-data}. Indeed, this data frame shows the \texttt{features} columns $z$ involving the initial state, the model's parameters and the control parameters while the last two columns shows the the boolean labels regarding the fulfillment of the tumor contraction  condition (defined by \eqref{TcontractionCond} with $\gamma_c=10^{-2}$) and the condition of the health indicator being safely above the threshold. The previous last columns show the normalized quantities of the three drugs being used [see \eqref{vj}] over the therapy duration (defined by \eqref{yH} in which $\rho=0.5$ is used). $T_f$ stands for the final values of the tumor. \\ \ \\
\tikz{
\node[rounded corners, fill=Blue!10, inner ysep=2mm, inner xsep=2mm] at(0,0){
\begin{minipage}{0.45\textwidth}
Fitting the binary classifiers using all the \texttt{features}	
\end{minipage} 
};
}
 \\ \ \\
The two classification problems corresponding to the derivation of the binary classifiers $F_T$ and $F_H$ (see item (7) of section \ref{sec-methodology}) are solved using the \texttt{RandomForestClassifier} module of the \texttt{scikit-learn} python Machine Learning library \cite{scikit-learn}. The 70000 lines of the learning data are split into learning data ($70\%$) and test data ($30\%$). Namely, the model is first derived using the training data before its prediction performance is tested on the never seen test data. The quality of the prediction of the identified model on the unseen test data is shown in Figure \ref{fig-prediction-all-binary}. 
\tikz{
\node[rounded corners, fill=Blue!10, inner ysep=2mm, inner xsep=2mm] at(0,0){
\begin{minipage}{0.45\textwidth}
Global sensitivity analysis of the \texttt{features} relevance	
\end{minipage} 
};
}
\\ \ \\
An appealing facility of the \texttt{RandomForestClassifier} model is the possibility, once a model \texttt{F} is fitted, to examine the order of importance of the \texttt{features} used to build the model. Indeed, the \texttt{F.feature\_importances\_} attribute represents an ordered list of \texttt{features} sorted by descending order of importance. \\ \ \\ Table \ref{tableimportantfeatures} shows the ordered important features used in both classifiers $F_T$ and $F_H$ while Figure \ref{fig-prediction-selected-binary} shows the resulting prediction performance when using only these selected \texttt{features} in fitting the classifiers. Note that while the number of features has been drastically reduced (from 43 to 5 and 7 respectively) the quality of the model is almost not impacted and even slightly improved regarding the prediction on unsuccessful health preservation since reducing the number of features reduces the risk of over-fitting and hence improve the robustness of extrapolation on unseen data. Note also that in both case, the most important feature is the state associated to the label to be predicted by the classifier, namely the initial value of the tumor for $F_T$ and the initial value of the circulating lymphocytes for $F_H$. It is also remarquable that only these two states are detrimental to the quality of the prediction while the values of the remaining state ($N$ and $L$) do not really matter. This will be very helpful in deriving the dashboards of indicators later on as the relevant state space is 2D while the ODE involves 6-states.
\begin{table}[H]
\begin{center}
\begin{tabular}{ll} \toprule
    {\bf Classifier} & {\bf Most important features}\\ \midrule
 $F_T$: Tumor contraction&  $x_1, d, x_4, \ell , a$\\
 $F_H$ Health condition&  $x_4, x_1, r, \beta_C, \kappa, k_C, T_\text{stop}$  \\ \bottomrule \\
\end{tabular}
\end{center} 
\caption{\color{Blue} The list of meaningful features for the fitting of the binary classifiers of tumor contraction $F_T$ and health preservation $F_C$.}
\label{tableimportantfeatures}
\end{table}

It is also important to note that among the 24 parameters involved in the model, only the following parameters seem to be important $d, \ell, a, k_C$. The first two parameters are involved in the definition of the quantity $D$ that concentrate the effect of the vaccine and the immunotherapy drugs on the tumor contraction. The parameter $a$ determine the natural expansion rate of the tumor while $k_C$ involved in the dynamics \eqref{dCdt} determine the impact of chemotherapy on the circulating lymphocytes which is obviously a key parameter. 
\\ \ \\
Another conclusion that comes out of Table \ref{tableimportantfeatures} is that the key parameters of the feedback design are: the rate of decrease $r$ of the tumor that stops the drug delivery according to \eqref{goodincrease}, the multiplicative security margin $\beta_C$ used in \eqref{betaCmuC} to express the health preservation constraint, the duty ratio $\kappa$ used to define the ratio of drug injection within the a period of therapy and finally, in a lesser extent\footnote{This is because removing this control parameter only marginally impact the precision of the prediction provided by the classifier $F_H$.}, the tumor size below which the chemotherapy is stopped.  
\ \\ \ \\
\tikz{
\node[rounded corners, fill=Blue!10, inner ysep=2mm, inner xsep=2mm] at(0,0){
\begin{minipage}{0.45\textwidth}
Fitting of the drug usage regressors using the selected relevant features.
\end{minipage} 
};
}
\\ \ \\
\begin{figure*}
\begin{center}
\includegraphics[width=\textwidth]{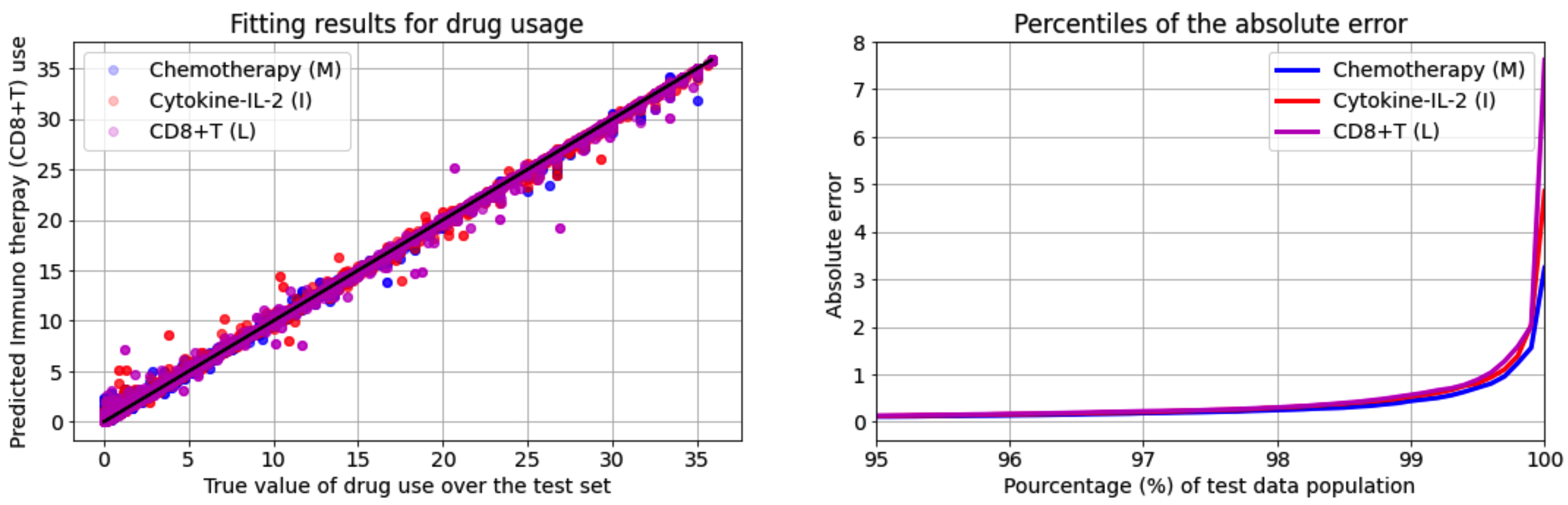} 
\end{center} 
\caption{\color{Blue} Results of the drug usage identification on the test dataset using the 10 most important \texttt{features} depicted on Table \ref{tableimportantfeatures} and using the \texttt{RandomForestRegressor} model of the \texttt{sciki-learn} library. {\bf (top)}: Chemotherapy, {\bf (Middle)} Immunotherapy and {\bf (bottom)} Vaccine.}\label{fig-id-drug-usage}
\end{figure*}
As explained in item (7) of section \ref{sec-methodology}, the training data can also be used to fit three regression models, namely $F_M$, $F_I$ and $F_L$ that capture the relationships that enables to predict the amount of drugs that would be used during the whole therapy duration for a given instance of initial state, model's parameter vector and control parameter vector. Here again, the \texttt{RandomForestRegressor} of the \texttt{scikit-learn} library is used in which only the 10 selected \texttt{feaatures} depicted on Table \ref{tableimportantfeatures} are used.\\ \ \\
Figure \ref{fig-id-drug-usage} shows the results regarding the prediction of the normalized drugs usage during the closed-loop simulation of the instances containing in the 21000 unseen test instances (30\% of the training data). The results clearly show very nice precision over a wide range of initial states, model's parameters and control parameters.
\ \\ \ \\
\tikz{
\node[rounded corners, fill=Blue!10, inner ysep=2mm, inner xsep=2mm] at(0,0){
\begin{minipage}{0.45\textwidth}
Using the fitted classifiers and regressors
\end{minipage} 
};
}
\\ \ \\
Throughout the remainder of this section, only the control parameters that belong to the set of \textit{important features} are optimized and hence involved in the so called decision variables. All the other parameters are fixed a priori to the values depicted in Table \ref{onminalvaluesapriori}.
\begin{table}[H]
\begin{center}
\begin{tabular}{lccr} \toprule
    {\bf Parameter} & {\bf value} & {\bf Parameter} & {\bf Value} \\ \midrule
$T_\text{max}$&  $10^8$&  $T_\text{th}$&  7 Days\\
$C_\text{min}$&  $3.125\times 10^{10}$&  $T_s$&  1 Day\\
$\bar v_M$&  $1$&  $\bar v_L$&  $10^7$\\
$\bar v_I$&  $10^4$\\
 $\tau$ &  $1$ hr&  $\gamma_c$& 0.01\\\bottomrule \\
\end{tabular}
\end{center} 
\caption{\color{Blue} Values of the parameters that are fixed a priori.}
\label{onminalvaluesapriori}
\end{table}
\ \\ \ \\
It is important to recall that the input (vector of \texttt{features}) of all the above mentioned classification and regression models contains the vector of model's parameters. But this vector is never available for a given patient. This is the reason why the above fitted models can only be used to compute the \textit{expectation} of the labels given a known statistical distribution of the vector of model's parameters. \\ \ \\
More precisely, for each values of the pair $$x_0^r:=(x_1^{(0)},x_4^{(0)})=(T_0,C_0)$$ of initial values of the tumor and the population size of the circulating lymphocytes, and a given value of the remaining reduced-dimensional vector of free control parameters: 
\begin{equation}
\theta:=(r, \beta_C, \kappa, T_\text{stop}) \in \Theta\label{defdetheta}
\end{equation} 
a closed-loop scenario for the pair $(x_0^r,p_\text{ctr}^r)$ is defined for each value of the remaining information needed to simulate the system, namely, the remaining values of the initial state vector and the vector of model's parameters. This information is gathered into the scenario vector $\omega$ defined by\footnote{Note that the initial concentration of drugs $x_5^{(0)}$ and $x_6^{(0)}$ are supposed equal to $0$ at the beginning of the therapy.}:
\begin{equation}
\omega := \begin{bmatrix}
x_2^{(0)}\cr  x_3^{(0)}\cr p_\text{model}
\end{bmatrix} \label{defdeomega}
\end{equation} 
Note that a cloud $\Omega$ containing $N$ values of this scenario vector can be drawn in accordance with the available knowledge on the dispersion of the model's parameters and state as (see Section \ref{SecgenerateLData}):
\begin{equation}
\Omega := \Bigl\{\omega^{(j)}\Bigr\}_{j=1}^N
\end{equation} 
Now since the arguments used as input of the fitted maps $F_\sigma()$, $\sigma\in \{T, H, M, I, L\}$ are included in the concatenation of the vectors $x_0^r, \theta$ and $\omega$, one disposes of the following five $x_0^r$-parametrized predictors (two classifiers for $\sigma\in \{T, H\}$ and three regressors for $\sigma\in \{M, i, L\}$:
\begin{equation}
F_\sigma^{(T_0, C_0)}(\theta, \omega)
\end{equation} 
each of which predicts a label that concerns the closed-loop behavior starting from a initial states {\bf sharing the initial conditions $\mathbf{(T_0, C_0)=x_0^r}$} while the remaining states are defined by the initial state components included in $\omega$ under the feedback control defined by the parameters $\theta$ and when the model parameters are defined by the component $p_\text{ctr}$ of $\omega$. In particular, the following prediction:
$$F_S^{(T_0,C_0)}(\theta, \omega):=\Bigl(F_T^{(T_0, C_0)}(\theta, \omega)\Bigr)\ \text{and}\  \Bigl(F_H^{(T_0, C_0)}(\theta, \omega)\Bigr)$$
indicates whether the therapy is successful under the conditions above or not. Now since $p_\text{model}$ is never known, the only effective computation would be the following estimation of probability of having successful therapy (when using the controller defined by $\theta$) among the scenarios included in $\Omega$, namely
\begin{equation}
\hat P^{(T_0, C_0)}_\text{success}(\theta):=\dfrac{1}{N}\sum_{j=1}^N \Bigl[F_S^{(T_0, C_0)}(\theta, \omega^{(j)})\Bigr]\label{defdeprob}
\end{equation} 
where the sum assumes that the value $1$ is associated to a $F_S$ being \texttt{true} while $0$ is associated to 
$F_S$ being \texttt{false}. 
Note however that the above expression is only an averaging-based approximation of the probability and holds true only asymptotically when $N\rightarrow\infty$. However, provided that one admits some probabilistically acceptable success level (say $1-\eta\approx 95\%$) and an associated confidence level (say $1-\delta\approx99.9\%$), then it is possible to determine the \textit{sufficient} number of samples, say $N(\eta,\delta,m)$ for a given number $m$ of accepted unsuccessful closed-loop scenarios such that if the number of un-successful scenarios is lower than $m$, namely:
\begin{equation}
\hat P^{(T_0, C_0)}_\text{success}(\theta)\ge 1-\dfrac{m}{N} \label{certifconstraint}
\end{equation} 
 then it can be certified with a confidence $1-\delta$ that the associated control parameter $\theta$ leads to successful closed-loop with probability $1-\eta$ for all the scenarios that share the same pair $(T_0, C_0)$. More precisely, denoting by $n_\Theta$ the cardinality of the discrete set $\Theta$ of control parameters, the above mentioned number of scenarios is given by \cite{alamo2009randomized}:
\begin{equation}
N\ge \dfrac{1}{\eta}\Biggl(m+\ln\left(\frac{n_\Theta}{\delta}\right)+\left((2m\ln\left(\frac{n_\Theta}{\delta}\right)\right)^{\frac{1}{2}}\Biggr)\label{formuleN}
\end{equation} 
Table \ref{lesNtab} shows the corresponding values of $N$ for $\delta=10^{-3}$ and $m=1$ for different values of the precision parameter $\eta$. 
\begin{table}
\begin{center}
\begin{tabular}{lcccc} \toprule
    {$n_\Theta$} & {$\eta=0.1$} & {$\eta=0.05$} & {$\eta=0.01$} & {$\eta=0.001$} \\ \midrule
    1  & 132 &       264 &      1317 &       13164 \\
    5  & 154 &       308  &     1536    &    15354\\
    10  & 163 &        326 &         1628    &    16280\\
    100  & 193 &       386 &       1930  &      19299\\
    1000  & 223 &       445 &       2225 &       22249\\
    10000  &  252  &     503   &    2515    &    25148\\
 \bottomrule
\end{tabular}
\end{center} 
\ \\
\caption{\color{Blue} Evolution of the sample size $N$ (number of trials needed to achieve the certification) as a function of the precision $\eta$ and the cardinality $n_\Theta$ of the design parameter set $\Theta$ (confidence parameter $\delta=10^{-3}$ and the number of unsuccessful scenarios $m=1$ are used)} \label{lesNtab} 
\end{table}
\ \\ \ \\
Regarding the set $\Theta$, the following definition is used in the forthcoming numerical investigations:
\begin{align}
\Theta:=\Bigl\{&(r, \beta_C, \kappa, T_\text{stop}) \in  \{0.1, 5, 10\}\times \{1.2, 1.5, 2\}\times\nonumber \\
&\times \{0.2, 0.5, 0.9\}\times \{10, 10, 1000\}\Bigr\}
\end{align} 
leading to a cardinality $n_\Theta=81$. Using Table \ref{lesNtab} together with $\eta=0.05$ leads to a number of scenarios $N=386$ which is used to produce the forthcoming results. 

\begin{figure}
\begin{center}
\includegraphics[width=0.48\textwidth]{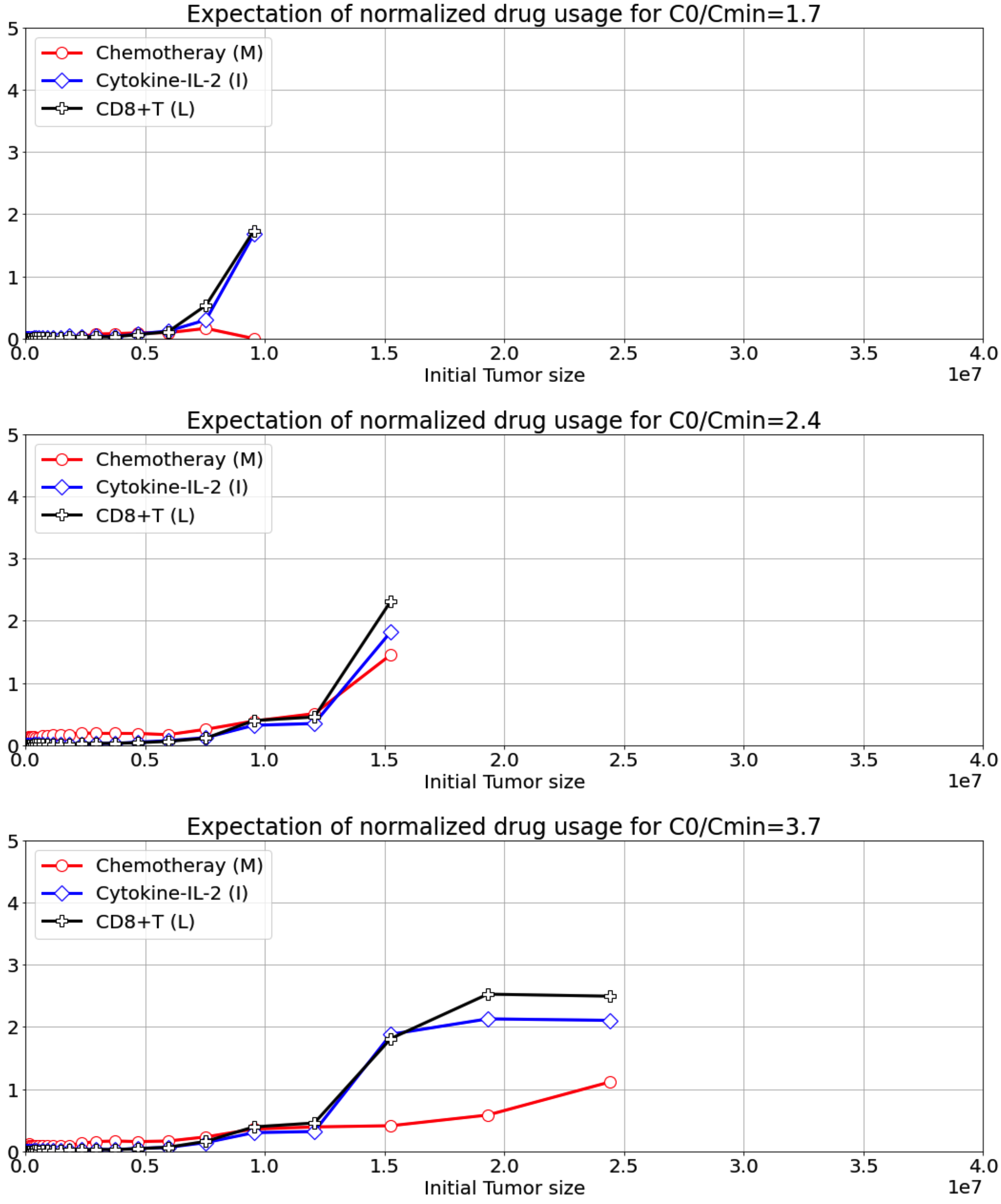} 
\end{center}
\caption{\color{Blue} Example of expected normalized quantities of the different drugs for a given value of the initial normalized size $(C(0)/C_\text{min})$ of circulating lymphocytes as a function of the initial tumor size. The abscissa range for which a successive therapy is possible depends on the initial value of $C(0)$. (level of uncertainties: $20\%)$}\label{fig-drugUsage-20} 	
\end{figure}

\begin{figure}
\begin{center}
\includegraphics[width=0.48\textwidth]{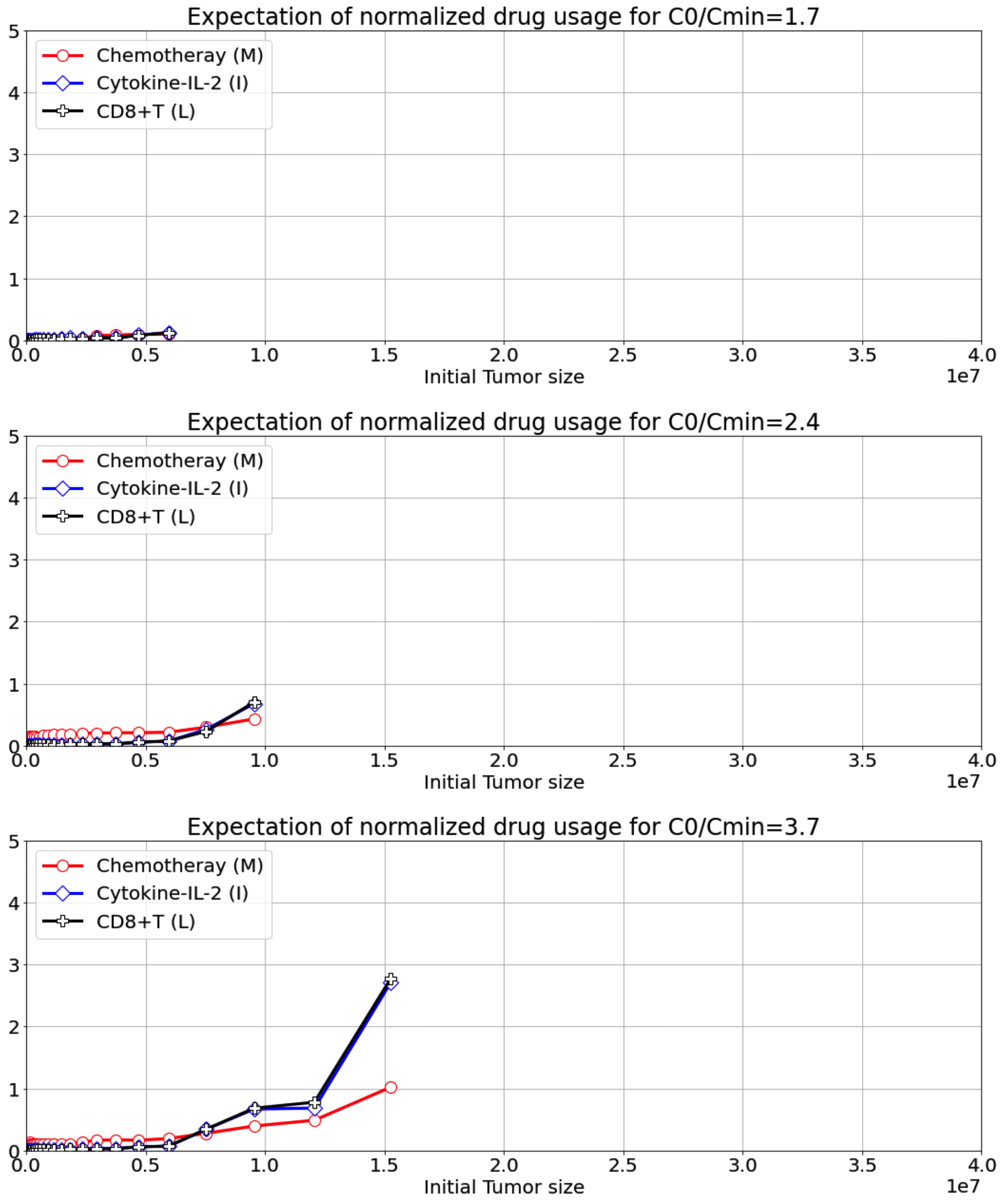} 
\end{center}
\caption{\color{Blue} Example of expected normalized quantities of the different drugs for a given value of the initial normalized size $(C(0)/C_\text{min})$ of circulating lymphocytes as a function of the initial tumor size. The abscissa range for which a successive therapy is possible depends on the initial value of $C(0)$. (level of uncertainties: $40\%)$}\label{fig-drugUsage-40} 	
\end{figure}

On the other hand, the expectation of the normalized quantities of drugs used under these circumstances can be approximated by the following expressions for $\sigma\in \{M, I, L\}$:
\begin{equation}
\hat Q_\sigma^{(T_0, C_0)}(\theta):=\dfrac{1}{N}\sum_{j=1}^N \Bigl[F_\sigma^{(T_0, C_0)}(\theta, \omega^{(j)})\Bigr]\label{defdeQsigma}
\end{equation} 
and assuming the relative prices\footnote{such that $\pi_M+\pi_I+\pi_L=1$.} $\pi_\sigma$ for $\sigma=M, I, L$, the expected cost can be computed according to:
\begin{equation}
\hat J^{(T_0, C_0)}(\theta):=\sum_{\sigma\in \{M,I,L\}}\pi_\sigma \hat Q_\sigma^{(T_0, C_0)}(\theta)
\end{equation} 
Having the above definitions at at hand, given a pair $(T_0, C_0)$ of initial tumor size and initial size of circulating lymphocytes' population, the best control parameter $\theta$ is determined by solving the following optimization problem:
\begin{equation}
\min_{\theta\in \Theta}\Bigl[\hat J^{(T_0, C_0)}(\theta)\Bigr]\quad \text{under}\quad \hat P^{(T_0, C_0)}_\text{success}(\theta)\ge 1-\dfrac{m}{N} \label{probtheta}
\end{equation} 
more precisely, when the problem admits a feasible solution (satisfying the constraint) then the cost function is minimized over the set of admissible values of $\theta$ and the pair $(T_0, C_0)$ is considered to lie in the success therapy region of the reduced state space. On the other hand, when there is no feasible solutions, the paire $(T_0, C_0)$ is declared to be outside the successful therapy region. 
\begin{figure}
\begin{center}
\includegraphics[width=0.48\textwidth]{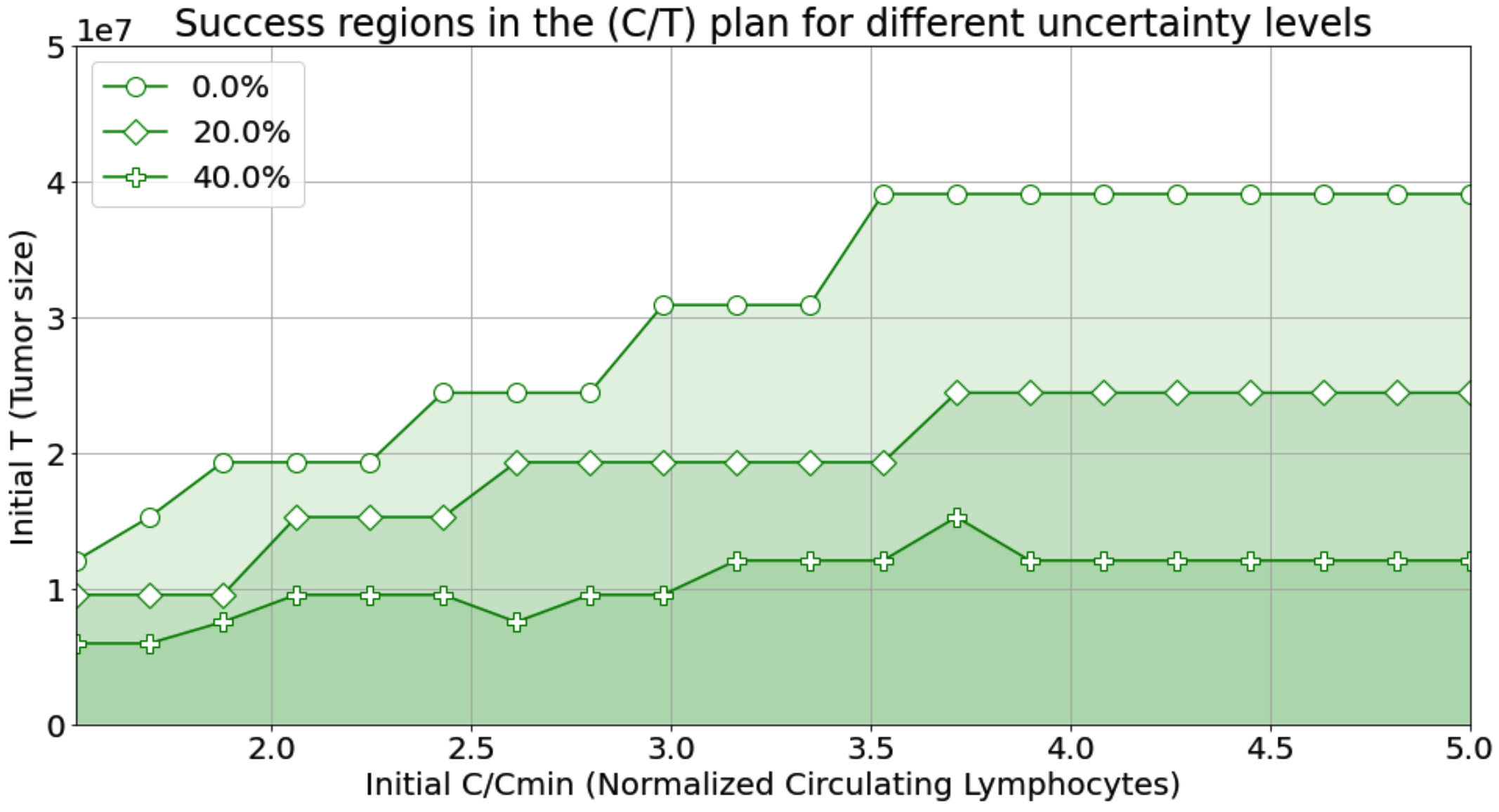} 
\end{center}
\caption{\color{Blue} Successful regions of therapy (in green) in the plan $(C-T)$ for different levels of parametric uncertainties and for the precision parameter $\eta=5\%$ and the confidence parameter $\delta=0.1\%$. }\label{fig-successRegions} 	
\end{figure}
Figure \ref{fig-successRegions} shows the successful regions characterization for different values of the uncertainty level regarding the model's parameters dispersion. As expected the size of the recoverable region decreases when the level of uncertainties increases.  
\begin{figure}
\begin{center}
\includegraphics[width=0.48\textwidth]{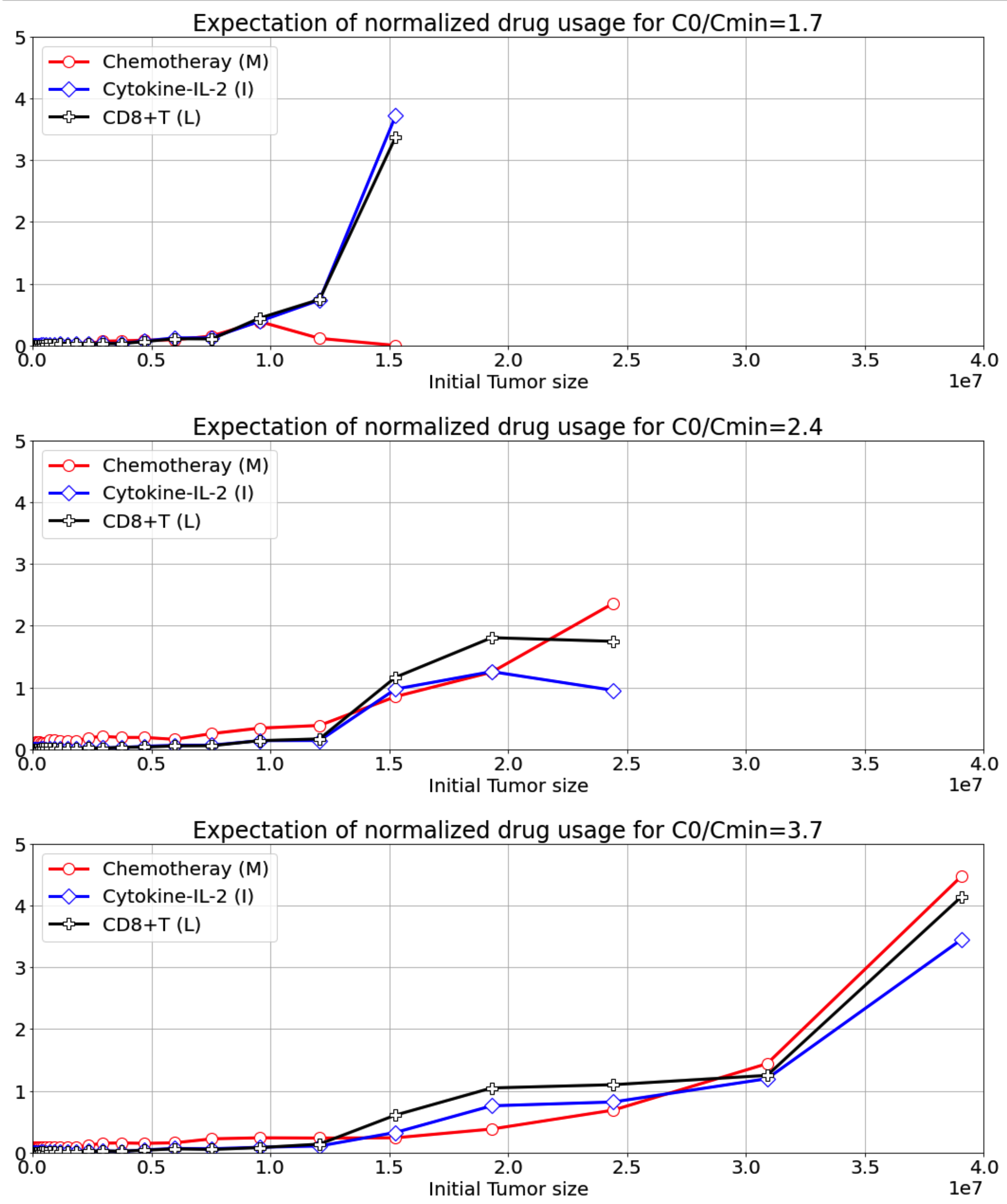} 
\end{center}
\caption{\color{Blue} Example of expected normalized quantities of the different drugs for a given value of the initial normalized size $(C(0)/C_\text{min})$ of circulating lymphocytes as a function of the initial tumor size. The abscissa range over  which a successful therapy is possible depends on the initial value of $C(0)$. (level of uncertainties: $0\%)$}\label{fig-drugUsage-00} 	
\end{figure}
Figures \ref{fig-drugUsage-00}, \ref{fig-drugUsage-20} and \ref{fig-drugUsage-40} show the evolution of the expected drugs usage during the therapy as a function of the initial tumor size $T_0$ for different values of the initial size of the population of circulating lymphocytes $C_0$

The results are computed only when a successful therapy is obtained which explains why the plots extends to higher values of the initial tumor size axis for smaller levels of parametric uncertainties. It can also be observed that when the uncertainty level increases, successful therapy are characterized by intensive use of immunotherapy and a moderate use of chemotherapy. The same observation can be made for very small valus of $C_0$ (upper sub-plots of all figures) where again, the use of chemotherapy is very moderate compared to the use of immunotherapy which is quite expected. 
\section{Conclusion}\label{sec-conc}
In this paper a general framework is proposed that enables to analyze the performance and the behavior of any explicit parameterized state feedback control strategy in the presence of high uncertainties on the model's parameters. It has also been shown that the use of Machine Learning modern tools enables to perform global sensitivity analysis which might be of great help to focus on the key quantities regarding both state components, model's parameters but also the control parameters. This enables to separate the choice of the set of parameters that can be quite rich in a first stage from the step where the meaningful parameters are found and selected for future development. 

The methodology has been validated on the specific case of combined therapy of cancer although it can be applied to a variety of drug scheduling in different diseases. 

It is important to note that the reduction of the state space dimension to those states that really matter in the final issue might render tractable the previous works involving non scalable approaches such as the one involved in \cite{alamir_robust_cancer:2014} and \cite{MoussaACC2020} which can therefore apply even to nominally high dimensional models. 
\ \\
\bibliography{bib_cancer}
\bibliographystyle{plain}
\end{document}